\documentclass[12pt]{article}

\pdfoutput=1

\usepackage{cite}
\usepackage{mathrsfs,amsmath,bm,multirow,color}

\evensidemargin=\oddsidemargin
\catcode`@=11

\let\@CITE=\@cite
\let\Cite=\cite
\def\@cite#1#2{{#1\if@tempswa , #2\fi}}

\def\cite#1{[\Cite{#1}]}
\def\textcite#1{Ref.~\Cite{#1}}

\catcode`@=12

\newcounter{nchoice}
\renewcommand{\thenchoice}{\Alph{nchoice}}
\def\chno#1{\refstepcounter{nchoice}\label{#1}{\rm\thenchoice}}

\def\Eqn#1{Eq.\ (\ref{#1})}
\def\Eqs#1#2{Eqs.\ (\ref{#1}) and (\ref{#2})}

\def\Sec#1{Sec.\,\ref{#1}}

\def\gclx{\ensuremath{{\rm SU(3)}_c \times {\rm SU(3)}_L \times {\rm
      U(1)}_X}}
\def\gcl{\ensuremath{{\rm SU(3)}_c \times {\rm SU(3)}_L}}
\def\gc3{\ensuremath{{\rm SU(3)}_c}}
\def\gl3{\ensuremath{{\rm SU(3)}_L}}
\def\gx1{\ensuremath{{\rm U(1)}_X}}
\def\gq1{\ensuremath{{\rm U(1)}_Q}}

\def\rep#1{\ensuremath{\mathit {#1}}}
\def\diag{\mathop{\rm diag}}

\title{\bf Generalized 331 models}

\author{\bf Pritibhajan Byakti\thanks{priti137@gmail.com}\\
  Pandit Deendayal Upadhyaya Adarsha Mahavidyalaya, \\ 
      Eraligool, Karimganj 788723, India 
   \\[5mm]
   \bf Palash B. Pal\thanks{palashbaran.pal@saha.ac.in} \\
   Physics Department, University of Calcutta,\\ 92 APC Road, Calcutta
   700009, India}

\date{}

\begin{document}

\maketitle

\begin{abstract}
  We examine different ways in which the standard model can be
  embedded into the \gclx\ gauge group.  We show that there exist 
  families of models characterized by a free parameter.  Only a few of
  these models, corresponding to specific values of the free
  parameter, have been studied so far.

\end{abstract}

\bigskip

\section{Introduction}
The standard model (SM) cannot predict the number of fermion
generations.  If one expands the gauge group to \gclx
\cite{Singer:1980sw, Valle:1983dk, Pisano:1991ee, Frampton:1992wt,
  Foot:1994ym}, then the number of generations come out to be 3, or
any multiple of 3.  In addition to this feature for which acted as a
prime motivation for constructing these models, the models can provide
an automatic solution to the strong CP problem \cite{Pal:1994ba} and
an interesting explanation of electric charge quantization
\cite{deSousaPires:1998jc, VanDong:2005ux}.  Because of these
attractive features, phenomenological implications of these models
have been extensively studied \cite{Buras:2012dp,
  CarcamoHernandez:2019iwh, Costantini:2019kmv,
  RobertsonSingh:2019jad, Dias:2020ryz, CarcamoHernandez:2020pxw,
  Lebbal:2020sqb, Hong:2020qxc}.  In addition, people have also
examined the Higgs potential \cite{Palcu:2019nld, Costantini:2020xrn},
possible candidates for cosmological dark matter \cite{Filippi:2005mt,
  deS.Pires:2007gi} and the possibilities of embedding of these models
\cite{Byakti:2013uya, Fonseca:2015aoa, Deppisch:2016jzl} into some
grand unified model.

For the sake of convenience, one often refers to the gauge group as
the 331 group, as has been done in the title of this article.  There
are different versions of 331 models, or really, different models with
the 331 gauge group.  Among the two early versions in which the number
of generations is predicted, right-handed neutrinos are absent in one
version \cite{Pisano:1991ee, Frampton:1992wt} and present in another
\cite{Singer:1980sw, Valle:1983dk, Foot:1994ym}.  The common feature
is the gauge group, the prediction of the number of generations, and
the fact that new fermions must be introduced in order to complete the
multiplets as well as to cancel gauge anomalies.  The difference lies
in the properties of these new fermions.  In addition, it was shown
later \cite{Deppisch:2016jzl} that if one chooses, one can also
construct models based on the same gauge group in which each
generation of fermions is independent in the sense that all anomalies
cancel within a single generation, and different generations are mere
copies of one another.  Of course, in this model the prediction of the
number of generations is lost, but the model is nonetheless viable.

The aim of this article is to examine whether other variants of the
model are possible.  We show that, for the case of non-identical
generations, the proposed models are specific examples of a class of
models characterized by a free parameter.  For the case of identical
fermion generations, however, the proposed model is unique under some
restrictive assumptions.  However, those assumptions are by no ways a
necessity, so they can be spared and more general models can be built.
We show that there are other kinds of models possible, which do not
resemble the earlier models in any way.

\section{Anomalies}\label{s:an}
So far as the color symmetry is concerned, this group is no different
from SM.  The quarks will be in color triplets and the leptons will be
in color singlets.  Since there will be new fields in these models,
let us say that anything that transforms like a triplet of \gc3\ will
be called quark, and any singlet of the color group will be called
lepton.

The \gl3\ part of the group requires obvious thoughts.  We assume that
all fermions transform either like the singlet or in the multiplet
whose dimension equals that of the fundamental representation, as is
the case in the SM.  This means that the representations under
\gl3\ are \rep1, \rep3 and \rep{3^*}.  We take all right-chiral fields
in \gl3\ singlets.  This is no loss of generality, because if we
encounter any right-chiral field that is not a singlet of \gl3, we can
disregard it and instead tabulate its complex conjugate which would
then be a left-chiral field.  With these ground rules, then, all
possible multiplets would fall under one of the categories listed in
Table~\ref{t:multiplets}.   It should be commented that
restriction to singlets and triplets does not follow from any
fundamental principle.  It is just an assumption.  There are models
where other representations are used \cite{Huong:2019vej}, but we will
not discuss them here.
\begin{table}
  \caption{Multiplets, showing their representations under the ${\rm
      SU(3)}_c \times {\rm SU(3)}_L$ part of the gauge group.  The ${\rm
      U(1)}_X$ values have been left unspecified.}\label{t:multiplets}

  $$
  \begin{array}{cccc}
    \hline
    \mbox{Type} & \mbox{Representation} & \mbox{Chirality} &
    \mbox{Range of $k$} \\ 
    \hline
    1 & (1,3,\alpha_k) & L & 1, \cdots, n_1  \\ 
    2 & (1,3^*,\beta_k) & L & 1, \cdots, n_2  \\
    3 & (3,3,\gamma_k) & L & 1, \cdots, n_3 \\ 
    4 & (3,3^*,\delta_k) & L & 1, \cdots, n_4 \\
    5 & (1,1,\xi_k) & R & 1, \cdots, n_5 \\
    6 & (3,1,\zeta_k) & R & 1, \cdots, n_6 \\
    \hline
  \end{array}
  $$
\end{table}

We now write down the constraints coming from gauge anomaly
cancellation.  While reading the ensuing equations, it has to be
remembered that the range of the sum on $k$ is in general different
for different lettered variables, as can be seen from
Table~\ref{t:multiplets}.
\begin{subequations}
  \label{anom}
\begin{align}
  [\gc3]^3~: && 3(n_3+n_4) - n_6 &= 0 \,,
  \label{ccc} \\
  [\gl3]^3~: && n_1 - n_2 + 3(n_3-n_4) &= 0 \,,
  \label{lll} \\
  [\gc3]^2 \times [\gx1]~: && 
  \sum_k \Big( 3 \gamma_k + 3 \delta_k - \zeta_k \Big) &= 0 \,,
  \label{ccx} \\
  [\gl3]^2 \times [\gx1]~: && \sum_k \Big( \alpha_k +
   \beta_k + 3  \gamma_k + 3  \delta_k \Big) &= 0
  \,, \label{llx} \\ 
  [\gx1]^3~: && \sum_k \Big( 3 \alpha_k^3 + 3 \beta_k^3 + 9
  \gamma_k^3 + 9 \delta_k^3 - \xi_k^3 - 3 \zeta_k^3 \Big) &= 0 \,. 
  \label{xxx}
\end{align}
In addition, gravitational anomaly cancellation would require the
condition
\begin{eqnarray}
  \sum_k \Big( 3 \alpha_k + 3 \beta_k + 9
   \gamma_k + 9 \delta_k - \xi_k - 3 \zeta_k \Big) &= 0 \,.
  \label{grav}
\end{eqnarray}
\end{subequations}

In the context of the SM gauge group, it was shown
\cite{Golowich:1990ki} that the anomaly cancellation equations, aided
by some reasonable assumptions like the presence of a mass of the
charged fermions, are enough to determine the charges of the fermion
fields.  This will not be the case in the context of the 331 models.
There are too many parameters, and one will need so many assumptions
to reach the goal that finally it would seem that the goals were
pre-determined and the assumptions were tailored to obtain them.  This
statement will be supported by the discussions in the rest of this
article, where we will see that there is enough freedom in the
construction which allows many models to be constructed.

Note that the $[\gc3]^3$ anomaly cancellation condition, \Eqn{ccc},
demands that the number of left-chiral triplets of color must equal
the number of right-chiral triplets.  That requirement is
automatically satisfied if we assume that all quarks are massive.
Similarly, \Eqn{lll} implies that the number of triplets and
antitriplets of \gl3\ must be equal.

Many of the other equations become trivial if we impose the
requirement that all charged particles must be massive.  In the mass
term, a left-chiral fermion field pairs with a right-chiral field.
Therefore, this requirement can be summarized by saying that all
non-trivial representations of the final unbroken group of $\gc3
\times \gq1$ should be vectorlike \cite{Golowich:1990ki}, the latter
factor being the gauge group of quantum electrodynamics.  Because the
electric charge generator is a linear combination of \gl3\ generators
and the \gx1\ generator, \Eqs{ccc} {ccx} together imply that the
$[\gc3]^2 \times \gq1$ anomalies must also cancel.  This will produce
a condition that is similar to \Eqn{ccx}, except that there will be
sums over the electric charges instead of the $X$ quantum numbers.
But that equation is automatically obeyed, because for each
left-chiral quark field there must be a right-chiral quark with the
same charge.  For the same reason, the quark terms in \Eqs{xxx} {grav}
must also vanish among themselves.  The same argument applies for the
lepton fields.  If there is a left-chiral field with a certain
non-zero charge, either there would be a type-5 right-chiral field of
the same charge present in the model, or a left-chiral field of
opposite charge must be present in the type-1 fields.  To summarize,
the anomaly conditions of \Eqs{xxx} {grav} need not be considered
separately if we arrange the right-chiral fields in such a way that
all left-chiral fields of nonzero charge can obtain masses.

\Eqs{lll} {llx} are then the important equations in the context of
model building.   Note that
all $n$'s must be non-negative integers by definition.  We can always
take
\begin{eqnarray}
  n_1 \geq n_2 
  \label{n1>n2}
\end{eqnarray}
by adjusting our definition of the \rep3 and \rep{3^*} representations
of \gl3.  \Eqn{lll} then implies that
\begin{eqnarray}
  n_4 \geq n_3 \,.
  \label{n4>n3}
\end{eqnarray}
We will also impose
\begin{eqnarray}
  n_1 \geq 1, \qquad n_4 \geq 1 \,,
  \label{n>1}
\end{eqnarray}
because otherwise the model would lack either leptons, or quarks, or
both.  Next, we note that \Eqn{lll} shows that $n_1-n_2$ must be a
multiple of 3.  We can explore how this equation can be satisfied,
taking small values of all $n$'s.  Here are some examples.
Corresponding to each pair of values of $n_3$ and $n_4$, we choose the
minimum values of $n_1$ and $n_2$, subject to the conditions of
\Eqs{n1>n2} {n>1}, that will satisfy \Eqn{lll}.
\begin{eqnarray}
  \begin{array}{ccccc}
    \mbox{Choice} & n_3 & n_4 & n_1 & n_2 \\  \hline
    \chno{ch01} & 0 & 1 & 3 & 0 \\
    \chno{ch11} & 1 & 1 & 1 & 1 \\
    \chno{ch12} & 1 & 2 & 3 & 0 
  \end{array}
  \label{n1-n2}
\end{eqnarray}
We will see that Choice~\ref{ch01} will give us the sequential model
\cite{Deppisch:2016jzl}, whereas Choice~\ref{ch12} will give us models
in which the generations are non-sequential, and  one needs all three
generations for all gauge anomalies to cancel \cite{Singer:1980sw,
  Valle:1983dk, Pisano:1991ee, Frampton:1992wt, Foot:1994ym}, which
can be called the entangled models.  An intermediate case,
corresponding to Choice~\ref{ch11}, provides a new kind of 331 model
which has not been explored so far.

Note that, so far we have not used \Eqn{llx} at all.  It will prove
useful when we will try to construct the models explicitly.

\section{Charge assignment}\label{s:ch}
The SM lepton doublet must lie within one of the $(1,3,\alpha_k)$
representations.  We can choose the triplet, for the first generation
of left-chiral lepton fields, to be
\begin{eqnarray}
  \left( \begin{array}{c} \nu_e \\ e \\ f 
  \end{array}  \right)_L \,,
  \label{nuef}
\end{eqnarray}
where $f_L$ is a lepton field that has to be added to complete the
triplet.  It might be a field that is already present in the SM, or
might be a new field --- the choice is left open for now.

The electric charge generator must be of the form
\begin{eqnarray}
  Q = T_L + X \,,
\end{eqnarray}
where $T_L$ is a diagonal generator of \gl3, and $X$ is the generator
of the \gx1\ part of the gauge group.  If one uses the standard
representations of SU(3) generators in which $T_3$ and $T_8$ are the
diagonal generators, then $T_L$ is a linear combination of those two.
For the \rep3 representation of \gl3, this means that the charges of
its components will be given by
\begin{eqnarray}
  Q_{(\rep3)} = \diag \Big( c_1 + x, c_2+x, c_3+x \Big) \,, 
  \label{Q3}
\end{eqnarray}
where $x$ is the \gx1\ quantum number, and
\begin{eqnarray}
  c_1 + c_2 + c_3 = 0
  \label{csum}
\end{eqnarray}
because they come from the \gl3\ generators which are traceless.  We
note that, since the representation of the generators in \rep3 and
\rep{3^*} are different, in the \rep{3^*} representation the electric
charge will be given by
\begin{eqnarray}
  Q_{(\rep{3^*})} = \diag \Big( {-}c_1 + x, -c_2+x, -c_3+x \Big) \,, 
  \label{Q3*}
\end{eqnarray}
with the same values of $c_1$, $c_2$ and $c_3$ that appear in
\Eqn{Q3}.

If the $X$ quantum number is taken to be $\alpha_1$ for the multiplet
shown in \Eqn{nuef}, the charges of the first two components would
imply the relations
\begin{eqnarray}
  c_1 + \alpha_1 = 0 \,, \qquad c_2 + \alpha_1 = -1 \,.
  \label{c1c2eq}
\end{eqnarray}
If we take the charge of the field $f$ to be $q$, \Eqn{csum} would
imply the relation
\begin{eqnarray}
  \alpha_1 = {q - 1 \over 3} \,.
  \label{alpha1}
\end{eqnarray}
So now we can find $c_1$, $c_2$ and $c_3$ in terms of $q$ by using
\Eqs{csum} {c1c2eq}, and thereby rewrite \Eqs{Q3}{Q3*} in the form
\begin{subequations}
  \label{Qq}
\begin{eqnarray}
  Q_{(\rep3)} &=& \diag \Big( {1-q \over 3} + x, \; - {2+q \over 3} +x, \; 
  {1+2q \over 3} +x \Big) \,,  
  \label{Q3q}  \\ 
  Q_{(\rep{3^*})} &=& \diag \Big( {q-1 \over 3} + x, \; {2+q \over 3} +x, \; 
  - {1+2q \over 3} +x \Big) \,.
  \label{Q3*q}
\end{eqnarray}
\end{subequations}

Let us now look at the quark sector.  We see from \Eqn{n1-n2} that
there must be at least one $(3,3^*)$ multiplet of \gcl.  Let us
denote its $X$ quantum number by $\delta_1$.  It must contain one of
the usual quark doublets, and they must occur in the first two
components of the \gcl\ representation.  However, note that 
\Eqs{c1c2eq} {Q3*q} tell us that, for a
\rep{3^*} representation of \gl3, the first component has lower charge
than the second one.  Therefore, we should identify the first
component as $d_L$ and the second one as $u_L$.  These identifications
imply
\begin{eqnarray}
  {q-1 \over 3} + \delta_1 = - \frac13 \,,
\end{eqnarray}
or
\begin{eqnarray}
  \delta_1 = -{q \over 3} \,.
  \label{delta1}
\end{eqnarray}
From this, we can find the charges of all three components of the
multiplet.  In particular, the third component will have a charge
$-\frac13-q$.  Exactly similarly, we can argue that, for a $(3,3)$
representation of \gcl, if the first two components are $u$-type and
$d$-type quarks, the third component would have a charge $\frac23+q$.
We will now use these results to construct the full models.  We will
not arrange our construction in the order of the choices given in
\Eqn{n1-n2}, Rather, we start with the possibility $n_2=0$, and will
later go up to $n_2=1$.

\section{Entangled models}\label{s:en}
To keep things as simple as possible, we will first assume, along with
$n_2=0$ as mentioned, that all $\alpha_k$'s are equal to $\alpha_1$
found in \Eqn{alpha1}, and similarly all $\gamma_k$'s and all
$\delta_k$'s are also equal.  We will denote this common value without
any subscripted index.  Then, \Eqn{llx} will imply
\begin{eqnarray}
  n_1 \alpha + 3n_3 \gamma + 3n_4 \delta = 0 \,.
\end{eqnarray}
The values of $\alpha$ and $\delta$ must be those given in
\Eqs{alpha1} {delta1}.  Then,
\begin{eqnarray}
  n_3 \gamma = - \frac13 n_1 \alpha - n_4 \delta = {(1-q) n_1
  + 3n_4q \over 9 } \,.
  \label{gamma}
\end{eqnarray}
This equation will be inconsistent if we take Choice~\ref{ch01} of
\Eqn{n1-n2}, where $n_3=0$.  Therefore, the next simplest solution is
Choice~\ref{ch12}, i.e., $n_1=3$, $n_3=1$ and $n_4=2$.  This will mean
that the three left-chiral quark doublets of SM appear in two
different types of representations of the 331 group.  The generations
are not identical, and all anomalies also do not cancel within a
single generation of fermions, which is why we call these models
`entangled'.

\Eqn{gamma} now gives 
\begin{eqnarray}
  \gamma = {q+1 \over 3} \,.  
\end{eqnarray}
Remarkably, this value coincides exactly with the value of the $X$
quantum number obtained from the requirement that a (3,3)
representation of \gcl\ contains the usual quark doublet.  Therefore
we find that the generalized 331 model will have the following
multiplet structure for the fermions:
\begin{eqnarray}
  \begin{array}{c@{\hspace{7mm}}c@{\hspace{7mm}}c@{\hspace{7mm}}l}
    \hline
    \mbox{\gclx} & \multirow2*{\mbox{Chirality}} & \mbox{Number of} &
    \mbox{electric} \\  
    \mbox{representation} && \mbox{copies} & \mbox{charges} \\ \hline
    \Big( 1,3, {q-1 \over 3} \Big) & L & 3 & 0,-1,q \\
    \Big( 3,3, {q+1 \over 3} \Big) & L & 1 & \frac23,-\frac13, \frac23+q \\
    \Big( 3,3^*, -{q \over 3} \Big) & L & 2 & -\frac13,\frac23,
    -\frac13-q \\ 
    \multicolumn4c{\mbox{and R-chiral fields to match their charges.}}
    \\ 
    \hline
  \end{array}
  \label{en.charges}
\end{eqnarray}
It can now be easily checked that  all anomaly cancellation conditions of
\Eqn{anom} are obeyed for any value of $q$ as long as the right chiral
fields contain the same charges as the left chiral fields, as argued
earlier in \Sec{s:an}.

Clearly, there are two distinguished values of $q$ that need to be
discussed.  One is $q=+1$, and the other is $q=0$.  For each of these
values, there arises the possibility that the new field can pair with
one of the two other fields of the same multiplet to form a mass term
for the fermion.  Interestingly, these values give the variants of the
331 models discussed in the literature: the $q=+1$ case gives the model
of Pisano, Pleitez and Frampton \cite{Pisano:1991ee, Frampton:1992wt},
whereas the model with $q=0$ was presented by Singer, Valle, Schechter,
Foot, Long and Tran \cite{Singer:1980sw, Valle:1983dk, Foot:1994ym}.
To the best of our knowledge, models with any other value of $q$
 have not been studied or proposed.

\section{Sequential models}\label{s:se}
In this case, we deal with the solution with $n_2=n_3=0$ that was
presented as Choice~\ref{ch01} in \Eqn{n1-n2}.  Since $n_1=3$ for
$n_4=1$, there will be three lepton multiplets corresponding to each
quark multiplet.  We can therefore use the quark multiplet to be the
marker of a fermion generation, and construct the field content of a
single generation, which can be repeated arbitrary number of times.
We can now write $\delta$ instead of $\delta_1$ since it is the only
parameter of its kind.  Clearly, we cannot take all $\alpha$'s to be
the same because, if we do that, then \Eqn{llx}, or equivalently
\Eqn{gamma}, is not obeyed with the values obtained in \Eqs{alpha1}
    {delta1}.

So at least two $\alpha_k$'s will be different.  Rather than
considering the possibility that they are all different, let us
consider the scenario where
\begin{eqnarray}
  \alpha_1 = \alpha_2 \neq \alpha_3 \,.
\end{eqnarray}
We now look at \Eqn{llx}.  With the values of $\alpha_1$ and $\delta$
from \Eqs{alpha1}{delta1}, we find
\begin{eqnarray}
  \alpha_3 = {2+q \over 3} \,.
\end{eqnarray}
We can now write the electric charges of all left-chiral fields in a
single generation:
\begin{eqnarray}
  \begin{array}{c@{\hspace{7mm}}c@{\hspace{7mm}}c@{\hspace{7mm}}l}
    \hline
    \mbox{\gclx} & \multirow2*{\mbox{Chirality}} & \mbox{Number of} & 
    \mbox{electric} \\  
    \mbox{representation} && \mbox{copies} & \mbox{charges} \\ \hline
    \Big( 1,3, {q-1 \over 3} \Big) & L & 2 & 0,-1,q \\
    \Big( 1,3, {q+2 \over 3} \Big) & L & 1 & 1,0, q+1 \\
    \Big( 3,3^*, -{q \over 3} \Big) & L & 1 & -\frac13,\frac23, -\frac13-q \\
    \multicolumn4c{\mbox{and R-chiral fields to match their charges.}}
    \\ 
    \hline
  \end{array}
  \label{seqmodel}
\end{eqnarray}
Here also, one can take any arbitrary value of $q$ and introduce a
right chiral field corresponding to every left chiral charged field
and obtain anomaly cancellation thereby.  But there is a more
economical possibility, which was found by Deppish, Hati, Patra,
Sarkar and Valle \cite{Deppisch:2016jzl}.  They entertained the
possibility that each charged left chiral field has a partner in the
form of a left chiral field of opposite charge.  Together, they can
form mass terms.  This means that there is no need for the singlets of
\gcl.  If this viewpoint is adopted, there is a unique value of $q$
that becomes acceptable.  This value comes from the fact that there
are two fields of charge $-1$ in the two copies of of the first kind
of multiplet shown in \Eqn{seqmodel}, and therefore we need two fields
with charge $+1$.  This can be achieved by putting $q=0$, so that both
fields of charge $+1$ can appear in the second kind of multiplet shown
in \Eqn{seqmodel}.  Setting $q=0$ and adding the right chiral quark
fields in \gcl\ singlets, one obtains a single generation of fermions
from \Eqn{seqmodel}.  Other generations are exact copies, so far as
the gauge transformation properties are concerned.  The model thus
obtained is then exactly the model suggested in
\textcite{Deppisch:2016jzl}, except for a difference in the
convention: what we call a triplet of \gl3\ was called an antitriplet
in \textcite{Deppisch:2016jzl}, and vice versa.

In order to exhaust all scenarios in which at least two $\alpha_k$'s
are equal, we should also examine the possibility
\begin{eqnarray}
  \alpha_1 \neq \alpha_2 = \alpha_3 \,.
\end{eqnarray}
Now, putting \Eqs{alpha1} {delta1} into \Eqn{llx}, we obtain
\begin{eqnarray}
  \alpha_2 = \alpha_3 = {1 + 2q \over 6} \,.
\end{eqnarray}
With these values of the $\alpha_k$'s, the following multiplets
result. 
\begin{eqnarray}
  \begin{array}{c@{\hspace{7mm}}c@{\hspace{7mm}}c@{\hspace{7mm}}l}
    \hline
    \mbox{\gclx} & \multirow2*{\mbox{Chirality}} & \mbox{Number of} & 
    \mbox{electric} \\  
    \mbox{representation} && \mbox{copies} & \mbox{charges} \\ \hline
    \Big( 1,3, {q-1 \over 3} \Big) & L & 1 & 0,-1,q \\
    \Big( 1,3, {1+2q \over 6} \Big) & L & 2 & \frac12, -\frac12,
    \frac12 +q \\ 
    \Big( 3,3^*, -{q \over 3} \Big) & L & 1 & -\frac13,\frac23, -\frac13-q \\
    \multicolumn4c{\mbox{and R-chiral fields to match their charges.}}
    \\ 
    \hline
  \end{array}
  \label{seqmodel2}
\end{eqnarray}
Note that these models necessarily contain fractionally charged
leptons.  Also note that there is no value of $q$ for which the
\gcl\ singlets can be avoided.  However, it should be emphasized that,
by adding suitable right chiral fields, one can definitely build a
viable model of this sort.

\section{Intermediate models}\label{s:in}
In the models discussed in \Sec{s:en}, anomalies do not cancel unless
we take all three generations of fermions together.  In the models of
\Sec{s:se}, anomalies cancel within a single generation.  We can now
explore the intermediate scenario in which one needs two generations
to cancel the gauge anomalies.  This means that we want $n_3+n_4=2$.
Among the integral solutions to this equation subject to \Eqn{n4>n3},
the solution $n_4=2, \ n_3=0$ merely duplicates the field content of
the sequential model.  Thus, we are left with the only other solution,
i.e., both $n_3$ and $n_4$ equals 1.  This means that $n_1=n_2$,
according to \Eqn{lll}.  If we take $n_1=n_2=0$ then there will be no
lepton in the model.  In order to accommodate leptons, we take the
next smallest solution, i.e., Choice~\ref{ch11} of \Eqn{n1-n2}.  In
order to accommodate the SM doublets of quarks in the \rep{(3,3)} and
the \rep{(3,3^*)} representations of \gcl, and also to accommodate an
SM lepton doublet in the \rep{(1,3)} representation, we need the
values of $\alpha$, $\gamma$ and $\delta$ as shown in
\Eqn{en.charges}.  \Eqn{llx} will then give the value of $\beta$.  We
summarize the information.
\begin{eqnarray}
  \begin{array}{c@{\hspace{7mm}}c@{\hspace{7mm}}c@{\hspace{7mm}}l}
    \hline
    \mbox{\gclx} & \multirow2*{\mbox{Chirality}} & \mbox{Number of} &
    \mbox{electric} \\  
    \mbox{representation} && \mbox{copies} & \mbox{charges} \\ \hline
    \Big( 1,3, {q-1 \over 3} \Big) & L & 1 & 0,-1,q \\
    \Big( 1,3^*, -{q+2 \over 3} \Big) & L & 1 & -1,0,-1-q \\
    \Big( 3,3, {q+1 \over 3} \Big) & L & 1 & \frac23,-\frac13, \frac23+q \\
    \Big( 3,3^*, -{q \over 3} \Big) & L & 1 & -\frac13,\frac23,
    -\frac13-q \\ 
    \multicolumn4c{\mbox{and R-chiral fields to match their charges.}}
    \\ 
    \hline
  \end{array}
  \label{in.charges}
\end{eqnarray}

The discussion does not imply that we can have only two generations
and no more if we follow this path.  It only means that all gauge
anomalies cancel among the two generations of fermions whose
specifications have been given in \Eqn{in.charges}.  In order to
confront phenomenology and include three generations of quarks, one
can always add a sequential generation as what has been discussed in
\Sec{s:se}.  This means that the complete model, with three
generations of fermions in it, will have
\begin{eqnarray}
  n_1 = 4, \qquad n_2 = 1, \qquad n_3 = 1, \qquad n_4 = 2.
\end{eqnarray}
\begin{table}[b]
  \caption{Comparison of field contents of the models described in
    the text.}\label{t:comparison}
  \begin{center}
    \begin{tabular}{lcccccl}
      \hline
      \multirow2*{Model} & \multirow2*{$n_1$} & \multirow2*{$n_2$} & \multirow2*{$n_3$} & \multirow2*{$n_4$} & Number of & Comment
      on \\
      &&&&& SM doublets & Economy \\
      \hline
      Entangled & 3 & 0 & 1 & 2 & 12 & Most economical \\
      Sequential & 9 & 0 & 0 & 3 & 18 & Least economical \\
      Intermediate & 4 & 1 & 1 & 2 & 14 & Intermediate \\
      \hline
    \end{tabular}
  \end{center}

\end{table}
\section{Comments}
We have identified many ways of extending the SM to a model based on
the gauge group \gclx.  For the entangled models where the model has a
prediction for the number of fermion generations, we show that there
is a whole family of models, characterized by a parameter $q$, which
are anomaly-free.  This family includes the models which have been
studied in detail in the literature \cite{Singer:1980sw, Valle:1983dk,
  Pisano:1991ee, Frampton:1992wt, Foot:1994ym}.  In contrast, there
are also sequential models, including one which has been studied in
some detail \cite{Deppisch:2016jzl}, in which generations are marked
by quark fields, and are copies of one another.   We have
shown that there can be an intermediate class of models in which there
is one sequential generation and two more generations which are
entangled through anomalies.  Many of these models have not been
studied at all.

It need not be said that one can make more complicated models with the
same gauge group.  It is possible to construct models with larger
number of fermion fields by adding any number of vectorlike fermions,
or gauge singlets, or multiple copies of the entire collection.  We
have only identified models which are minimal corresponding to some
set of assumptions.

These minimal models can be compared on the basis of their field
content.  In Table\,\ref{t:comparison}, we summarize the total number
of triplets and antitriplets of \gl3\ in each kind of model.  Each of
these representations contain a doublet of the standard electroweak
model.  We see that the completely entangled models have no SM doublet
other than the ones which are already present in the three generations
of the SM.  The sequential models need a lot of new doublets, whereas
the intermediate models are intermediate in this aspect as well.  The
number of singlets of \gl3\ are different in the individual models of
each kind, and are not included in the table.

\paragraph*{Acknowledgements~:} The research of PBP was supported
by the SERB grant EMR/2017/001434 of the Government of India.

\bibliographystyle{unsrt}
\bibliography{gen331.bib}

\end{document}